\documentclass[aps,prl,reprint,superscriptaddress,amsmath,amssymb,longbibliography,floatfix]{revtex4-2}
\usepackage{graphicx,bm}
\usepackage{tikz}
\usetikzlibrary{decorations.pathmorphing,decorations.markings,arrows.meta}
\usepackage[colorlinks=true,linkcolor=blue,citecolor=blue,urlcolor=blue,breaklinks=true]{hyperref}
\usepackage{xcolor,capt-of,needspace}
\graphicspath{{figures/}}
\newcommand{\Lc}{\Lambda_c^+\bar\Lambda_c^-}
\newcommand{\ee}{e^+e^-}
\newcommand{\GeV}{\,\mathrm{GeV}}
\newcommand{\MeV}{\,\mathrm{MeV}}
\newcommand{\fb}{\,\mathrm{fb}}
\newcommand{\pb}{\,\mathrm{pb}}
\newcommand{\dd}{\mathrm{d}}
\hypersetup{pdftitle={ISR Reconstruction of Radiative C=+1 Charmonia as the Origin of the Belle-BESIII Line-Shape Puzzle}}
\makeatletter
\def\ps@sm{\def\@oddhead{\hfill\normalfont\thepage}\let\@evenhead\@oddhead\def\@oddfoot{}\let\@evenfoot\@oddfoot}
\makeatother
\begin{document}
\title{ISR Reconstruction of Radiative \texorpdfstring{$C=+1$}{C=+1} Charmonia as the Origin of the Belle--BESIII Line-Shape Puzzle}
\author{Cheng-Xi Liu}
\email{liuchx2025@lzu.edu.cn}
\affiliation{School of Physical Science and Technology, Lanzhou University, Lanzhou 730000, China}
\affiliation{Lanzhou Center for Theoretical Physics, Key Laboratory of Theoretical Physics of Gansu Province, Key Laboratory of Quantum Theory and Applications of MoE, Gansu Provincial Research Center for Basic Disciplines of Quantum Physics, Lanzhou University, Lanzhou 730000, China}
\affiliation{MoE Frontiers Science Center for Rare Isotopes, Lanzhou University, Lanzhou 730000, China}
\affiliation{Research Center for Hadron and CSR Physics, Lanzhou University and Institute of Modern Physics of CAS, Lanzhou 730000, China}

\author{Tian-Le Gao}
\email{gaotl2023@lzu.edu.cn}
\affiliation{School of Physical Science and Technology, Lanzhou University, Lanzhou 730000, China}
\affiliation{Lanzhou Center for Theoretical Physics, Key Laboratory of Theoretical Physics of Gansu Province, Key Laboratory of Quantum Theory and Applications of MoE, Gansu Provincial Research Center for Basic Disciplines of Quantum Physics, Lanzhou University, Lanzhou 730000, China}
\affiliation{MoE Frontiers Science Center for Rare Isotopes, Lanzhou University, Lanzhou 730000, China}
\affiliation{Research Center for Hadron and CSR Physics, Lanzhou University and Institute of Modern Physics of CAS, Lanzhou 730000, China}

\author{Cheng-Ping Shen}
\email{shencp@fudan.edu.cn}
\affiliation{Key Laboratory of Nuclear Physics and Ion-beam Application (MOE) and Institute of Modern Physics, Fudan University, Shanghai 200443, China}

\author{Xiang Liu}
\email{xiangliu@lzu.edu.cn}
\affiliation{School of Physical Science and Technology, Lanzhou University, Lanzhou 730000, China}
\affiliation{Lanzhou Center for Theoretical Physics, Key Laboratory of Theoretical Physics of Gansu Province, Key Laboratory of Quantum Theory and Applications of MoE, Gansu Provincial Research Center for Basic Disciplines of Quantum Physics, Lanzhou University, Lanzhou 730000, China}
\affiliation{MoE Frontiers Science Center for Rare Isotopes, Lanzhou University, Lanzhou 730000, China}
\affiliation{Research Center for Hadron and CSR Physics, Lanzhou University and Institute of Modern Physics of CAS, Lanzhou 730000, China}

\begin{abstract}
A longstanding disagreement persists between Belle and BESIII over the line shapes of $\ee\to\Lc$, $D_sD_{s1}(2536)$, and $D_sD_{s2}^{*}(2573)$ near $4.63\GeV$. Belle observes enhancements through initial state radiation that are absent or much weaker at BESIII. We propose that the extra Belle yield arises from radiative production of a $C=+1$ charmonium. Luminosity normalization increases this contribution to the extracted cross sections, so tens of fb can explain the enhancements. Belle II can test this mechanism through full reconstruction and searches for missing charmonia in open charm modes.
\end{abstract}

\maketitle

\noindent\emph{Introduction.---}The observation of charmonium-like $XYZ$ states has opened a window onto hadron spectroscopy and deepened our understanding of the nonperturbative behavior of the strong interaction. Among these states, the $Y$ states---charmonium-like states produced in electron--positron annihilation---constitute the so-called $Y$ problem~\cite{WangCharmoniumFamily2019,WangHigherCharm2020}, which has motivated a variety of theoretical efforts~\cite{ChenNonresonantY4260,ChenNonresonantY4360,ChenFanoY2016,ChenInterferenceY2018,HeMissingPsi4S2014,ChenMissingPsi4S2015,ChenMissingPsi4S2016,WangPsi4500,WangOpenCharmLevels2024,PengEtaJPsi2024,ManSDMixing2025}; see Refs.~\cite{ChenNewHadrons2023,OpenCharmReview,BaiUnquenched2027} for recent reviews.

A long-standing puzzle in this context concerns the line shape of the $\Lambda_c\bar{\Lambda}_c$ invariant mass spectrum. In 2008, the Belle Collaboration reported a higher $Y$ state, $Y(4630)$, in $e^+e^-\to \Lambda_c\bar{\Lambda}_c$~\cite{BelleLc}. With the accumulation of experimental data, the BESIII Collaboration searched for $Y(4630)$ through the same process~\cite{BESLc18,BESLc23} but found no enhancement structure in this energy range. This discrepancy constitutes a persistent Belle--BESIII difference in the $\Lambda_c\bar{\Lambda}_c$ line shape. Despite serious cross-checks by both collaborations, no explanation has been found.

Remarkably, this phenomenon is not unique. With the BESIII and Belle measurements of $e^+e^-\to D_s D_{s1}(2536)$ and $e^+e^-\to D_s D_{s2}^*(2573)$~\cite{BelleDs1,BelleDs2,BESDsJ}, the same Belle--BESIII difference emerges in the line shapes of the corresponding invariant mass spectra. This suggests an underlying dynamical mechanism common to these channels, the identification of which is a timely task in this area.

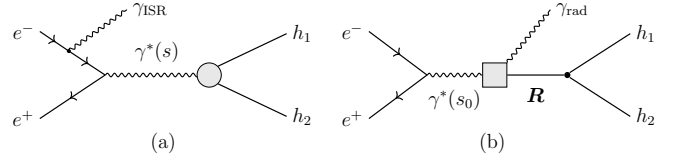
\begin{figure}[!t]
  \centering

  \resizebox{\columnwidth}{!}{%
  \begin{tikzpicture}[x=1cm,y=.9cm,font=\large,
    photon/.style={decorate,decoration={snake,amplitude=1.05pt,segment length=4.2pt},line width=.65pt},
    fermion/.style={line width=.7pt,postaction={decorate},decoration={markings,mark=at position .54 with {\arrow{>}}}},
    hadron/.style={line width=.7pt},
    blob/.style={circle,draw,fill=gray!18,minimum size=5mm,inner sep=0pt},
    decay/.style={circle,fill=black,draw=none,minimum size=3.6pt,inner sep=0pt},
    production/.style={rectangle,draw,fill=gray!18,minimum width=5mm,minimum height=5mm,inner sep=0pt}]
    \begin{scope}
      \draw[fermion] (0,1)--(.6,.55);
      \node[left] at (0,1) {$e^-$};
      \draw[fermion] (.6,.55)--(1.35,0);
      \draw[fermion] (1.35,0)--(0,-1) node[left] {$e^+$};
      \draw[photon] (.6,.55)--(1.8,1.5) node[right] {$\gamma_{\rm ISR}$};
      \fill (.6,.55) circle (1.2pt);
      \draw[photon] (1.35,0)--(3.5,0) node[midway,above=4pt] {$\gamma^*(s)$};
      \node[blob] (vectorvertex) at (3.5,0) {};
      \draw[hadron] (vectorvertex.north east)--(5.1,.95) node[right] {$h_1$};
      \draw[hadron] (vectorvertex.south east)--(5.1,-.95) node[right] {$h_2$};
      \node at (2.55,-1.55) {(a)};
    \end{scope}
    \begin{scope}[xshift=6.8cm]
      \draw[fermion] (0,1) node[left] {$e^-$} -- (1.2,0);
      \draw[fermion] (1.2,0)--(0,-1) node[left] {$e^+$};
      \node[production] (radiativevertex) at (2.6,0) {};
      \node[decay] (decayvertex) at (4.1,0) {};
      \draw[photon] (1.2,0)--(radiativevertex.west) node[midway,below=5pt] {$\gamma^*(s_0)$};
      \draw[photon] (radiativevertex.north east)--(3.75,1.5) node[right] {$\gamma_{\rm rad}$};
      \draw[hadron] (radiativevertex.east)--(decayvertex.west) node[midway,below=4pt,font=\large] {$\bm{R}$};
      \draw[hadron] (decayvertex.north east)--(5.4,.95) node[right] {$h_1$};
      \draw[hadron] (decayvertex.south east)--(5.4,-.95) node[right] {$h_2$};
      \node at (2.55,-1.55) {(b)};
    \end{scope}
  \end{tikzpicture}}

  \caption{(a) ISR production of the hadron pair $h_1h_2$ at Belle. The circle is the transition $\gamma^*(s)\to h_1h_2$ also measured by BESIII. (b) Radiative production of $R$ at Belle's energy, followed by $R\to h_1h_2$. The square is $\gamma^*(s_0)\to\gamma_{\rm rad}R$; the filled dot is the decay. Here $s$ is the squared invariant mass of $h_1h_2$ and $s_0$ the squared collision energy.}
  \label{fig:production}
\end{figure}

Motivated by the strong coupling of higher $\eta_c$ states to $\Lambda_c^+\bar{\Lambda}_c^-$~\cite{LiuEta2026}, we conjecture that the $\Lambda_c\bar{\Lambda}_c$ invariant mass spectrum measured by Belle contains a higher $\eta_c$ signal. The key point is the reconstruction method used in initial state radiation (ISR) analyses. Full reconstruction requires reconstructing both hadrons from their decay products, as in Belle's $D\bar D$ analysis~\cite{BelleDD}. Their measured momenta allow the photon momentum to be inferred and its angular distribution to be determined without photon detection. In the $\Lambda_c^+\bar\Lambda_c^-$ channel, Belle uses partial reconstruction: one $\Lambda_c$ and an energetic photon are reconstructed, and the other baryon is identified through recoil constraints~\cite{BelleLc}. Radiative events can satisfy the photon energy and recoil mass requirements and enter Belle's $\gamma\Lambda_c^+\bar\Lambda_c^-$ sample.

In this Letter, we extend this line of reasoning to the $D_s D_{s1}(2536)$ and $D_s D_{s2}^*(2573)$ channels. We show that the inclusion of higher charmonium contributions naturally accounts for the BESIII--Belle difference in the line shapes of these invariant mass spectra: higher charmonia lying above the $D_s D_{s1}(2536)$ and $D_s D_{s2}^*(2573)$ thresholds can couple to these channels and produce additional resonant contributions that may appear in the Belle measurements but be absent in those of BESIII. Normalization by the small effective ISR luminosity increases their contribution to Belle's extracted cross sections, allowing radiative products of tens of fb to generate visible enhancements and account for the observed line shape differences. The Supplemental Material (SM)~\cite{SupplementalMaterial} gives the fitting methods and results of the spectroscopy calculations.\nocite{ChenVMD,Song:2015D,Song:2015Ds}

\noindent\emph{Production mechanism.---}Figure~\ref{fig:production} compares the two topologies yielding a photon $\gamma$ and two hadrons $h_1,h_2$. Here $\sqrt{s_0}$ is the collision energy and $m=\sqrt{s}$ the hadron pair invariant mass. Panel (a) shows ISR at Belle; after photon emission, the virtual photon produces the hadron pair through the same transition measured by BESIII:
\begin{equation}
 \ee\to\gamma_{\rm ISR}\,\gamma^*(s)\to\gamma_{\rm ISR}\,h_1h_2.
 \label{eq:isr}
\end{equation}
Panel (b) is radiative production of a charmonium resonance $R$ at the full Belle energy,
\begin{equation}
 \ee\to\gamma^*(s_0)\to\gamma_{\rm rad}R,\qquad R\to h_1h_2.
 \label{eq:radiative}
\end{equation}
Both photons in panel (b) have $C=-1$, so charge conjugation fixes $C(R)=+1$. Radiative production allows $\eta_c$ states with $J^{PC}=0^{-+}$ and $\chi_{cJ}$ states with $J^{PC}=J^{++}$, $J=0,1,2$; these labels include all states with the corresponding quantum numbers. At $\sqrt{s_0}=10.58\GeV$, a resonance near $4.63\GeV$ recoils against a photon whose energy in the collision center of mass is
\begin{equation}
 E_\gamma^*=\frac{s_0-s}{2\sqrt{s_0}}\simeq4.28\GeV.
 \label{eq:photonenergy}
\end{equation}
In a BESIII direct scan at the same hadron pair mass, the collision energy is used to produce the pair, leaving no energy for a hard photon.

The ISR and radiative processes have the same $\gamma h_1h_2$ final state and can satisfy the same recoil requirements. In a fully reconstructed analysis, the squared mass recoiling against the two hadrons, $M_{\rm rec}^2(h_1h_2)$, is required to be close to zero, as expected for one missing photon. The three Belle analyses considered here use partial reconstruction: one hadron and an energetic photon are reconstructed, and the other hadron is identified through recoil constraints. Requiring full reconstruction of only one charm hadron improves efficiency, especially when the accessible decay branching fractions are small~\cite{BelleLc,BelleDDstar}. For $\Lc$, Belle reconstructs one $\Lambda_c^+$, requires the photon energy to exceed $3.5\GeV$, and identifies the recoiling $\bar\Lambda_c^-$ by $M_{\rm rec}(\Lambda_c^+\gamma)$ together with an antiproton from its decay~\cite{BelleLc}. The two $D_sD_{sJ}$ analyses impose the same kind of recoil constraint and require a photon above $3\GeV$ in the collision frame~\cite{BelleDs1,BelleDs2}. Equation~\eqref{eq:photonenergy} shows that an ISR photon and a radiative photon have the same energy at the same pair mass, so the cut on the photon energy and the recoil mass do not separate them.

In addition, full reconstruction allows the photon momentum and direction to be inferred from the two hadron momenta through energy and momentum conservation, without requiring photon detection. Belle's $D\bar D$ analysis used this method and observed the characteristic ISR peaks along the beams~\cite{BelleDD}. In the partially reconstructed samples considered here, the photon polar angle $\theta_\gamma^*$ relative to the electron beam in the collision center of mass is measured directly from the detected photon. Its angular coverage is limited by the calorimeter acceptance~\cite{BelleDetector}, excluding the directions closest to the beams where ISR is most strongly peaked. The predicted photon distribution for radiative production has no sharp peaks along the beams~\cite{ChungLeeYu}. Within the central window $|\cos\theta_\gamma^*|<0.9$ used in several Belle ISR analyses~\cite{BelleDD,BelleDDtwo,BelleDDstarPi,BelleDsGround}, the ISR~\cite{Binner} and radiative angular shapes are similar~\cite{SupplementalMaterial}. The restricted acceptance and present statistics therefore limit their separation. We add their yields because interference cancels when final states and their charge conjugates are combined with equal weight~\cite{Binner}.

Belle normalizes the event yield in each hadron mass bin using the effective ISR luminosity,
\begin{equation}
 \mathcal L_{{\rm ISR},j}=\mathcal L_B
 \int_{m_j^-}^{m_j^+}\frac{2m}{s_0}
 W\!\left(s_0,1-\frac{m^2}{s_0}\right)\,\dd m,
 \label{eq:isrluminosity}
\end{equation}
where $\mathcal L_B$ is the collected integrated luminosity, $m_j^\pm$ are the edges of mass bin $j$, and $W$ is the ISR radiator, the probability density in the photon energy fraction $1-m^2/s_0$~\cite{BelleDs1,BelleISRLuminosity}. In ISR, an incoming lepton emits a photon before annihilation, lowering the effective annihilation energy to the invariant mass $m$ of the hadron pair. Selecting a narrow hadron mass bin restricts this photon to a narrow energy interval. The small emission probability in that interval is the integral in Eq.~\eqref{eq:isrluminosity}, so $\mathcal L_{{\rm ISR},j}\ll\mathcal L_B$.

Radiative production in Eq.~\eqref{eq:radiative} proceeds through annihilation at the full collision energy, with the photon produced together with $R$. No preceding ISR emission is needed to lower the annihilation energy. For channel $i$ and mass bin $j$, $\sigma^{\rm rad}_{ij}$ is the radiative product cross section integrated over the bin and normalized with the full Belle luminosity $\mathcal L_B$. If these events enter the ISR sample, their contribution to Belle's reported cross section is extracted using the smaller effective ISR luminosity $\mathcal L_{{\rm ISR},j}$; we denote this contribution by $\sigma_{R,ij}$. 

The radiative product is the $\ee\to\gamma R$ production cross section multiplied by $\mathcal B(R\to h_1h_2)$ and any daughter branching fractions retained in the measured observable. The published baryon cross section is corrected for the $\Lambda_c$ reconstruction and inclusive $\bar\Lambda_c^-\to\bar pX$ branching fractions~\cite{BelleLc}, while the meson products retain the $D_{sJ}$ decay branching fractions~\cite{BelleDs1,BelleDs2}. The corresponding reconstruction efficiencies are $\epsilon^{\rm ISR}_{ij}$ and $\epsilon^{\rm rad}_{ij}$. We multiply Belle's dressed meson cross sections and uncertainties by the vacuum polarization correction factor $0.94$ to express them in the Born convention used for the BESIII vector baselines~\cite{BelleDs1,BelleDs2}. We quote $\sigma_{R,ij}$ and $\sigma^{\rm rad}_{ij}$ with the same Born normalization. Normalizing the radiative contribution with the two luminosities gives
\begin{equation}
 \sigma_{R,ij}=\frac{\mathcal L_B}{\mathcal L_{{\rm ISR},j}}
 \frac{\epsilon^{\rm rad}_{ij}}{\epsilon^{\rm ISR}_{ij}}\,\sigma^{\rm rad}_{ij}.
 \label{eq:mapping}
\end{equation}
We set $\epsilon^{\rm rad}_{ij}/\epsilon^{\rm ISR}_{ij}=1$. Counting radiative events as ISR therefore multiplies their cross section by $\mathcal L_B/\mathcal L_{{\rm ISR},j}$. Belle reports $\mathcal L_B=695\fb^{-1}$ for $\Lc$ and $\mathcal L_B=921.9\fb^{-1}$ for each meson channel~\cite{BelleLc,BelleDs1,BelleDs2}. In $D_sD_{s1}$, the bin $4.61<m<4.63\GeV$ has $\mathcal L_{{\rm ISR},j}=90.6\pb^{-1}=0.0906\fb^{-1}$, efficiency $0.132\%$ including the branching fractions of the reconstructed $D_s^+$ decay modes, and a fitted signal yield of $21.4$ events~\cite{BelleDs1}. If these events were radiatively produced, their contribution in the Born normalization adopted for the meson fits would be
\begin{equation}
 \sigma_{R,ij}=\frac{0.94\times21.4}{90.6\times0.00132}\,\mathrm{pb}\simeq168.2\pb.
 \label{eq:isrexample}
\end{equation}
The same yield corresponds to a radiative product of
\begin{equation}
 \sigma^{\rm rad}_{ij}=\frac{0.94\times21.4}{921.9\times0.00132}\,\mathrm{fb}\simeq16.5\fb.
 \label{eq:radexample}
\end{equation}
With $\mathcal L_B/\mathcal L_{{\rm ISR},j}=1.02\times10^4$, a radiative product of $16.5\fb$ in this bin would contribute $168.2\pb$ to the spectrum in the same Born normalization.

\noindent\emph{Fits and candidate analysis.---}We describe the Belle invariant mass spectra as the sum of the vector and radiative contributions,
\begin{equation}
 \sigma(s)=\sigma_V(s)+\sigma_R(s),
 \label{eq:model}
\end{equation}
where $\sigma$ is the total Born cross section fitted to Belle's ISR spectrum, $\sigma_V$ the vector contribution common to Belle ISR and BESIII direct scans, and $\sigma_R$ the extra contribution extracted when radiative events are normalized with the ISR effective luminosity.

For $\Lc$ we use the cross section of Ref.~\cite{ChenVMDOriginal}, including the $\psi(4500)$, $\psi(4660)$, $\psi(4790)$, and $\psi(4900)$ contributions and their interference. Keeping its resonance masses, width parameters, and fitted form factor parameters at their published values, we compare the baseline with the sixteen BESIII points from $4.575$ to $4.951\GeV$ and obtain $\chi^2/\mathrm{dof}=8.19/16$~\cite{BESLc18,BESLc23}.

For the meson channels, we adopt BESIII's published total fits containing two interfering vector Breit--Wigner amplitudes near $4.6\GeV$ and $4.7$--$4.8\GeV$. BESIII reports $\chi^2/\mathrm{dof}=4.0/8$ for $D_sD_{s1}$ and $6.2/7$ for $D_sD_{s2}^{*}$~\cite{BESDsJ}. These vector baselines are fixed as background in the Belle fits.

We describe $\sigma_R$ with a relativistic Breit--Wigner and the phase space for $\gamma h_1h_2$,
\begin{equation}
 \sigma_R(s)=\Phi^{(3)}(s)\left|C\,f_L(q)\,\mathrm{RBW}(s)\right|^2.
 \label{eq:resonance}
\end{equation}
Here, $\Phi^{(3)}(s)$ is the phase space density for $\gamma h_1h_2$ as a function of the hadron pair mass $m=\sqrt{s}$. Radiative production of such highly excited charmonia may involve complex unquenched effects~\cite{BaiUnquenched2027}, making the production strengths difficult to calculate. We therefore determine the overall normalization $C\geq0$ from a fit to each Belle spectrum. In the fitted line shape, we treat the smooth luminosity conversion factor as constant near the resonance and absorb it into $C$. For resonance mass $M_R$ and width $\Gamma_R$ at $s=M_R^2$, the relativistic Breit--Wigner function is $\mathrm{RBW}(s)=M_R\Gamma_R/[M_R^2-s-iM_R\Gamma_R(s)]$. Its width depends on the hadron momentum $q(s)$ in the $h_1h_2$ rest frame and relative orbital angular momentum $L$ as
\begin{equation}
 \Gamma_R(s)=\Gamma_R\left[\frac{q(s)}{q_R}\right]^{2L+1}
 \frac{M_R}{\sqrt{s}}\left[\frac{B_L(q)}{B_L(q_R)}\right]^2,
 \label{eq:runningwidth}
\end{equation}
with $q_R=q(M_R^2)$ and Blatt--Weisskopf barrier factors $B_L$~\cite{vonHippelQuigg1972}. We use a barrier radius $r=3\GeV^{-1}$ \cite{LHCbBarrier2015}. The decay factor is $f_L(q)=(q/q_R)^LB_L(q)/B_L(q_R)$~\cite{LHCbBarrier2015}, with $f_0=1$. For a massless photon, standard kinematics~\cite{PDG2025} gives
\begin{equation}
 \Phi^{(3)}(s)=\frac{q(s)}{32\pi^3}\left(1-\frac{s}{s_0}\right).
 \label{eq:phasespace}
\end{equation}
Below threshold, $\sigma_R=0$. We use $L=0$ for $\Lc$. For the allowed decays of $\chi_{cJ}$ states, we use $L=1$; for $\eta_c$ states decaying into $D_sD_{s2}^{*}$, we use $L=2$. The bin averages $\bar\sigma_{R,ij}$ in the Born convention used in the fits give the radiative product over the mass window $[m_a,m_b]$:
\begin{equation}
 \sigma_i^{\rm rad}[m_a,m_b]\simeq\sum_{j\subset[m_a,m_b]}
 \bar\sigma_{R,ij}\,\frac{\mathcal L_{{\rm ISR},j}}{\mathcal L_B}
 \frac{\epsilon^{\rm ISR}_{ij}}{\epsilon^{\rm rad}_{ij}}.
 \label{eq:conversion}
\end{equation}
For the luminosity conversion, we treat all Belle data as taken at $\sqrt{s_0}=10.58\GeV$.

The $\Lc$ fit uses sixteen Belle bins of $20\MeV$ from $4.58$ to $4.90\GeV$~\cite{BelleLc}. The meson fits cover $4.51$--$4.95\GeV$ for $D_sD_{s1}$ and $4.54$--$4.94\GeV$ for $D_sD_{s2}^{*}$~\cite{BelleDs1,BelleDs2}; their uncertainty treatment is given in the SM~\cite{SupplementalMaterial}. For $\Lc$, the error bars in Belle's published figure show statistical uncertainties only. We include the $10\%$ experimental systematic of Ref.~\cite{BelleLc} in each bin, and an estimated common $4.9\%$ uncertainty based on recent $\Lambda_c$ branching fraction measurements~\cite{BESLcBF2026,BESLcN2023}. The data term with common normalization $n$ is
\begin{equation}
 \chi^2_{\rm data}(n)=\sum_j\frac{[y_j-n\bar\sigma_j]^2}{u_j^2}.
 \label{eq:chi}
\end{equation}
Here, $j$ labels the fitted mass bins, $y_j$ is the measured cross section, and $\bar\sigma_j$ is the total prediction averaged over bin $j$. The factor $n$ accounts for the common branching fraction uncertainty. The uncertainty $u_j$ combines the statistical and $10\%$ experimental systematic errors in quadrature. The full $\chi^2$ also includes the normalization constraint given in Ref.~\cite{SupplementalMaterial}. Uncertainties on fitted parameters are the $\Delta\chi^2=1$ profile intervals~\cite{PDG2025}.

\begin{figure*}[t]
 \centering
 \includegraphics[width=\textwidth]{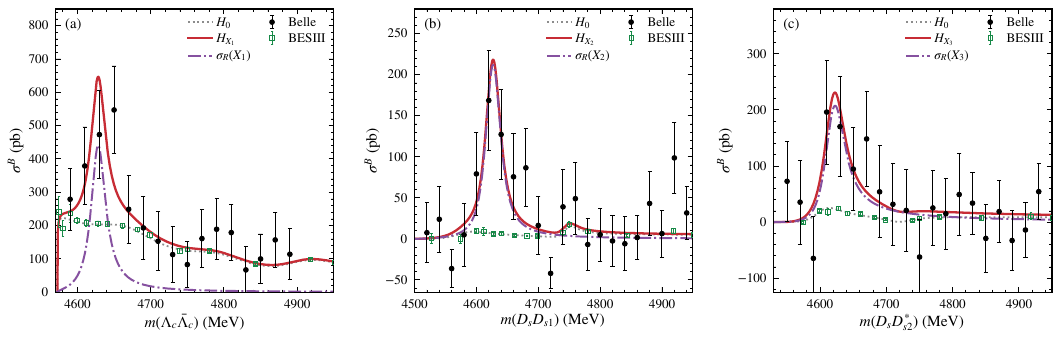}
 \caption{Fits to the (a) $\Lc$, (b) $D_sD_{s1}$, and (c) $D_sD_{s2}^{*}$ spectra. Black circles and green open squares show Belle and BESIII data. Grey dotted curves show the fixed vector baselines $H_0=\sigma_V$. Red solid and purple dash-dotted curves show the total fits $H_{X_i}$ and radiative contributions $\sigma_R(X_i)$, including the fitted normalization. The labels $X_1$, $X_2$, and $X_3$ correspond to (a)--(c). The mass and width of $X_1$ are fixed; those of $X_2$ and $X_3$ are fitted. Error bars include statistical and systematic uncertainties.}
 \label{fig:fits}
\end{figure*}

We denote the additional resonances in $\Lc$, $D_sD_{s1}(2536)$, and $D_sD_{s2}^{*}(2573)$ by $X_1$, $X_2$, and $X_3$, respectively. Figure~\ref{fig:fits} shows the fits to these three spectra. Our previous $B^0\to K_S^0\Lc$ analysis~\cite{LiuEta2026} studied an $\eta_c(6S)$ contribution to the $\Lc$ spectrum through an $S$-wave decay. We therefore consider this state as a candidate for $X_1$ and adopt the fixed input $(M_1,\Gamma_1)=(4629,24.2)\MeV$ from that study. As shown in Fig.~\ref{fig:fits}(a), this radiative contribution can describe Belle's enhancement near $4.63\GeV$ above the fixed vector baseline. The fit reduces $\chi^2/\mathrm{dof}$ from $17.12/16$ for the vector baseline alone to $6.81/15$, with a local significance of $3.2\sigma$. We also allow the mass and width to vary, obtaining $M_1=4640.1^{+6.0}_{-13.0}\MeV$ and $\Gamma_1=28.7^{+33.8}_{-23.2}\MeV$, with $\chi^2/\mathrm{dof}=4.67/13$. Both fixed inputs lie within the corresponding individual $1\sigma$ intervals of this best fit~\cite{SupplementalMaterial}.

For $X_2$, the fit gives $M_2=4627.3^{+6.7}_{-7.0}\MeV$ and $\Gamma_2=23.9^{+16.2}_{-15.2}\MeV$, lowering $\chi^2/\mathrm{dof}$ from $35.72/22$ for the vector baseline to $21.61/19$. For $X_3$, the fit gives $M_3=4625.1^{+20.3}_{-9.5}\MeV$ and $\Gamma_3=35.9^{+66.7}_{-28.3}\MeV$, with $\chi^2/\mathrm{dof}=8.87/17$, compared with $17.12/20$ for the baseline alone. With their masses and widths fixed at the fitted values, the local significances are $3.8\sigma$ for $X_2$ and $2.9\sigma$ for $X_3$. The fitted radiative products are $\sigma_1^{\rm rad}=76.9^{+25.2}_{-24.5}\fb$, $\sigma_2^{\rm rad}=42.9^{+14.1}_{-13.4}\fb$, and $\sigma_3^{\rm rad}=60.7^{+29.6}_{-26.6}\fb$~\cite{SupplementalMaterial}. These contributions of tens of fb can produce the Belle enhancements through the ISR luminosity normalization. The $\eta_c(6S)$ inputs used for $X_1$ lie within the corresponding individual $1\sigma$ intervals of the free $X_3$ fit. Fixing $(M_3,\Gamma_3)=(4629,24.2)\MeV$ gives $\chi^2/\mathrm{dof}=9.69/19$~\cite{SupplementalMaterial}, so an $\eta_c(6S)$ contribution can also describe this spectrum. Since $\eta_c\to D_sD_{s1}$ is forbidden, we consider $\chi_{c0}$, $\chi_{c1}$, and $\chi_{c2}$ states with allowed $P$-wave decays to this channel as candidates for $X_2$.

To examine possible charmonium assignments, we calculate the spectrum with a screened potential and the strong decays with the quark pair-creation (QPC) model~\cite{GodfreyIsgur,WangHigherCharm2020,Micu,LeYaouanc,WangCharmoniumFamily2019,WangPsi4500}, retaining the parameters of our previous study~\cite{LiuEta2026}. The central calculated masses of the $\chi_{c0}$, $\chi_{c1}$, and $\chi_{c2}$ states are~\cite{SupplementalMaterial}

\begin{align*}
 M[\chi_{cJ}(5P)]&=(4552.4,4563.1,4572.4)\MeV,\\
 M[\chi_{cJ}(6P)]&=(4678.1,4685.9,4693.0)\MeV.
\end{align*}

The fitted $X_2$ mass lies between the central predictions for these two multiplets. Orbital mixing has been considered in studies of vector charmonia~\cite{WangCharmoniumFamily2019,WangPsi4500}. Among the $\chi_{cJ}(6P)$ states, only $\chi_{c2}$ has an $F$-wave partner with the same $J^{PC}=2^{++}$. Possible mixing of $\chi_{c2}(6P)$ with $\chi_{c2}(5F)$, whose calculated mass is $4726.4^{+33.7}_{-34.7}\MeV$, could lower the $\chi_{c2}(6P)$ mass into the fitted range. We discuss this mixing possibility for a $\chi_{c2}$ interpretation of $X_2$ in the SM~\cite{SupplementalMaterial}. In addition, the possibly mixed $\chi_{c2}(6P)$ candidate may also contribute to the $D_sD_{s2}^{*}$ channel. However, the calculated \(\eta_c(6S)\) partical width into this channel is larger than those of the mixed tensor candidates, supporting \(\eta_c(6S)\) as a candidate for \(X_3\).

\noindent\emph{Summary.---}We have proposed radiative production of higher charmonia as a possible common origin of the differences between the Belle and BESIII line shapes. Radiatively produced $C=+1$ charmonia can pass Belle's partial reconstruction requirements, and normalization by the effective ISR luminosity increases their contribution to the extracted cross sections. The baryonic and $D_sD_{s2}^{*}$ enhancements can be described by $\eta_c(6S)$ contributions, while a $\chi_{c2}$ candidate can describe $D_sD_{s1}$. Possible $P$--$F$ mixing offers an interpretation of the tensor candidate in the latter channel. The spectrum and decay properties of highly excited charmonia remain incompletely understood, and further studies are needed to identify the contributing states.

With larger samples, comparing photon angular distributions in the enhancement region and neighboring mass intervals would help test whether the excess contains a radiative component. At Belle II, full reconstruction of both hadrons would allow the recoil photon direction to be inferred from their momenta, helping distinguish ISR from radiative production. Hadron decay angles would also provide information on the resonance spin. For example, angular correlations involving $D_{s2}^{*}\to DK$ in the $D_sD_{s2}^{*}$ channel could help distinguish the $\eta_c(6S)$ and $\chi_{c2}$ assignments.

Similar differences in the line shapes may also occur in other open charm channels. Many Belle points in $D^{*+}D^{*-}$ between $4.2$ and $4.4\GeV$ lie above the BESIII measurements, although the two data sets remain consistent within uncertainties~\cite{OpenCharmReview}. In this region, well above the $D^{*+}D^{*-}$ threshold, the large vector baseline means that an additional radiative contribution may produce only a modest relative increase in the total cross section. Channels opening closer to the resonance region, such as $D^*\bar D_2^*(2460)$, may therefore be more promising for observing a similar effect. Close collaboration between theorists and experimentalists will help test this mechanism and search for similar resonance structures in other channels.

\begin{acknowledgments}
This work is supported by the Natural Science Foundation of Gansu Province (No. 26RCKA012 and No. 25JRRA799), the National Natural Science Foundation of China under Grants No. 12335001, No. 12247101, and No. 12135005, the 111 Center under Grant No. B20063, the National Key R\&D Program of China under Contract No. 2024YFA1610503, the Fundamental Research Funds for the Central Universities (lzujbky-2023-stlt01), and Lanzhou City High-Level Talent Funding. T.-L. G. is supported by the Gansu Province Postgraduate Innovation Star Program No. 2026CXZX-042. The authors used ChatGPT for language editing and for assistance with the numerical code, reviewed the output, and take responsibility for the content.
\end{acknowledgments}

\emph{Data availability.} The cross sections come from Refs.~\cite{BelleLc,BESLc18,BESLc23,BelleDs1,BelleDs2,BESDsJ} and their supplemental material. The Belle $D_sD_{s2}^{*}$ points are digitized from the published figure.
\par

\clearpage
\setcounter{secnumdepth}{2}
\renewcommand{\thesection}{\Roman{section}}
\renewcommand{\theequation}{S\arabic{equation}}
\renewcommand{\thefigure}{S\arabic{figure}}
\renewcommand{\thetable}{SM-\Roman{table}}
\setcounter{equation}{0}
\setcounter{figure}{0}
\setcounter{table}{0}
\pagestyle{sm}
\thispagestyle{sm}
\raggedbottom

\begin{center}
{\Large\bfseries Supplemental Material}\\[6pt]
{\large ISR Reconstruction of Radiative \texorpdfstring{$C=+1$}{C=+1} Charmonia as the Origin of the Belle--BESIII Line-Shape Puzzle}
\end{center}

This Supplemental Material gives the fitting methods and results of the spectroscopy calculations.

\section{Fitting methods}
\label{sm:fits}

\subsection{Photon angular distributions}
\label{sm:angles}

The photon polar angle $\theta_\gamma^*$ is measured relative to the electron beam in the collision center of mass. The radiative distributions follow the calculation at leading order in Ref.~\cite{ChungLeeYu}, and the ISR distribution follows Ref.~\cite{Binner}. Each curve is averaged over bins of width $0.05$ in $\cos\theta_\gamma^*$ and normalized to unit area within $|\cos\theta_\gamma^*|<0.9$.

The full angular range is shown to display the ISR peaks along the beams. When both hadrons are reconstructed, their momenta determine the recoil photon direction, allowing the ISR peaks along the beams to be studied without detecting the photon~\cite{BelleDD}. In the partially reconstructed samples considered here, the photon must be detected, and the calorimeter acceptance excludes the directions closest to the beams.

We use $|\cos\theta_\gamma^*|<0.9$ as the central comparison interval, following several Belle ISR analyses~\cite{BelleDD,BelleDDtwo,BelleDDstarPi,BelleDsGround}. Figure~\ref{fig:smangles} shows that the ISR and radiative distributions have similar shapes in this interval, where the strongest ISR peaks are excluded. The restricted angular coverage and present statistics therefore limit separation of the two processes.

\noindent\begin{minipage}{\columnwidth}
\centering
\includegraphics[width=.8\columnwidth]{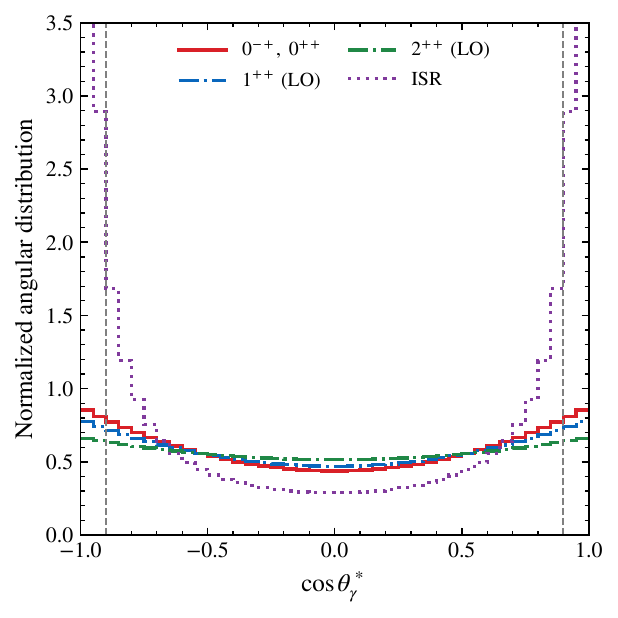}
\setlength{\abovecaptionskip}{3pt}
\captionof{figure}{Theoretical photon angular distributions at $\sqrt{s_0}=10.58\GeV$ and hadron pair mass $m=4.63\GeV$. Steps show averages over $0.05$ bins. Each distribution has unit area within $|\cos\theta_\gamma^*|<0.9$, marked by the vertical gray dashed lines.}
\label{fig:smangles}
\end{minipage}

\subsection{\texorpdfstring{Data and fitting method}{Data and fitting method}}

The fits use $20\MeV$ bins: $16$ bins over $4.58$--$4.90\GeV$ for $\Lc$, $22$ over $4.51$--$4.95\GeV$ for $D_sD_{s1}(2536)$, and $20$ over $4.54$--$4.94\GeV$ for $D_sD_{s2}^{*}(2573)$~\cite{BelleLc,BelleDs1,BelleDs2}. Belle's collected integrated luminosity $\mathcal L_B$ is $695\fb^{-1}$ for the baryonic channel and $921.9\fb^{-1}$ for each meson channel~\cite{BelleLc,BelleDs1,BelleDs2}. The central values of the published cross sections are retained. Channels related by charge conjugation are included throughout. The $D_sD_{s2}^{*}$ points are digitized from the published figure~\cite{BelleDs2}.

For $i=1,2,3$, corresponding to $\Lc$, $D_sD_{s1}$, and $D_sD_{s2}^{*}$, respectively, $\sigma_i^{\rm rad}$ denotes the radiative product including the decay branching fractions $\mathcal B(R_i\to h_1h_2)\mathcal B_{{\rm sub},i}$. Here $R_i$ is the additional resonance and $h_1h_2$ the hadron pair in channel $i$. These products use the same Born normalization as the fitted spectra.

For $\Lc$, $\mathcal B_{{\rm sub},1}=1$, because Belle's published cross section has been corrected for the $\Lambda_c$ reconstruction and inclusive $\bar\Lambda_c^-\to\bar pX$ branching fractions~\cite{BelleLc}. The measured products in the meson channels include $\mathcal B_{{\rm sub},2}=\mathcal B(D_{s1}^-\to\bar D^{*0}K^-)$ for $D_sD_{s1}$ and $\mathcal B_{{\rm sub},3}=\mathcal B(D_{s2}^{*-}\to\bar D^0K^-)$ for $D_sD_{s2}^{*}$, while the branching fractions of the reconstructed $D_s$ decay modes are included in the efficiencies~\cite{BelleDs1,BelleDs2}.

We average the fitted Born contribution $\sigma_{R,i}$ over each experimental mass bin and denote the result by $\bar\sigma_{R,ij}$. The radiative product over the mass window $[m_a,m_b]$ is then
\begin{equation}
\sigma_i^{\rm rad}[m_a,m_b]\simeq\sum_{j\subset[m_a,m_b]}\bar\sigma_{R,ij}\frac{\mathcal L_{{\rm ISR},j}}{\mathcal L_B}\frac{\epsilon^{\rm ISR}_{ij}}{\epsilon^{\rm rad}_{ij}}.
\label{eq:smconversion}
\end{equation}
The ISR luminosity $\mathcal L_{{\rm ISR},j}$ is defined in the Letter~\cite{BelleISRLuminosity,BelleDs1,BelleDs2}. We set $\epsilon^{\rm rad}_{ij}/\epsilon^{\rm ISR}_{ij}=1$, so each bin contribution is multiplied by $\mathcal L_{{\rm ISR},j}/\mathcal L_B$ before summing over the mass window. For the luminosity conversion, we treat all Belle data as taken at $\sqrt{s_0}=10.58\GeV$.

We fit the bin averages of the total cross section described in the Letter, with the vector baselines fixed as background~\cite{ChenVMDOriginal,ChenVMD,BESLc18,BESLc23,BESDsJ}. For each channel $i$, the data contribution to $\chi_i^2$ is
\begin{equation}
\chi_{{\rm data},i}^2(\bm a_i,n_i)=\sum_j\frac{[y_{ij}-n_i\bar\sigma_{ij}(\bm a_i)]^2}{u_{ij}^2}.
\label{eq:smchi}
\end{equation}
Here $y_{ij}$ is the measured cross section, $\bar\sigma_{ij}$ is the average total prediction in bin $j$, and $\bm a_i$ denotes the resonance parameters varied in the fit. The uncertainty $u_{ij}$ combines the statistical error and systematic errors that are independent between bins.  A single factor $n_i$ multiplies the total prediction in every bin and is fitted together with the resonance parameters. The total $\chi_i^2$ also includes the Gaussian constraint $(n_i-1)^2/\delta_{n,i}^2$, where $\delta_{n,i}$ is the relative common normalization uncertainty. The upper and lower $1\sigma$ limits for each parameter are the values at which $\chi_i^2$ increases by $1$ from its minimum. The other free parameters, including $n_i$, are refitted when determining these limits~\cite{PDG2025}.\par

\subsection{\texorpdfstring{Experimental uncertainties}{Experimental uncertainties}}
\label{sm:uncertainties}

Belle's original $\Lc$ figure gives statistical errors only~\cite{BelleLc}. Its total systematic uncertainty of about $42\%$ is dominated by branching fraction uncertainties. We retain the $10\%$ experimental contribution in the error for each point and estimate the common branching fraction uncertainty from recent BESIII measurements.

BESIII measures $\mathcal B(\Lambda_c^+\to pK^-\pi^+)=(6.61\pm0.11\pm0.12)\%$~\cite{BESLcBF2026} and $\mathcal B(\bar\Lambda_c^-\to\bar nX)=(32.4\pm0.7\pm1.5)\%$~\cite{BESLcN2023}. Including subsequent antihyperon decays, $\mathcal B(\bar\Lambda_c^-\to\bar pX)=1-\mathcal B(\bar\Lambda_c^-\to\bar nX)$ gives $(67.6\pm0.7\pm1.5)\%$. Using the precision of $pK^-\pi^+$ as a reference for the reconstructed $\Lambda_c$ modes, we conservatively add the two relative errors, $r_{\rm BF}\simeq2.5\%+2.4\%=4.9\%$, to estimate the common normalization uncertainty.

For $D_sD_{s1}$, we estimate the common normalization uncertainty as $4.9\%$ based on Belle's efficiency variation, retaining the remaining systematic errors for each point~\cite{BelleDs1}. For $D_sD_{s2}^{*}$, we treat the full $10.9\%$ systematic uncertainty as a common normalization uncertainty and use the statistical error for each point~\cite{BelleDs2}.

\subsection{\texorpdfstring{Fit results}{Fit results}}
\label{sm:results}

\begin{figure*}[!htbp]
\centering
\includegraphics[width=.96\textwidth]{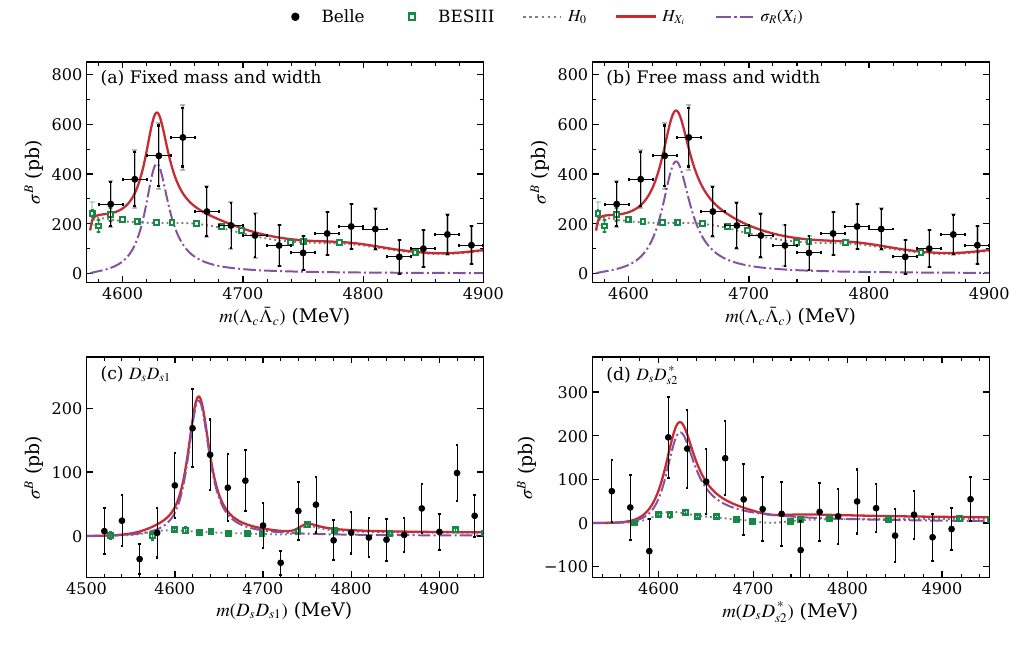}
\caption{Fits to Belle data (black circles) for $\ee\to\Lc$ in (a,b), $\ee\to D_sD_{s1}$ in (c), and $\ee\to D_sD_{s2}^{*}$ in (d); green open squares show BESIII data. The resonance mass and width are fixed in (a) and fitted in (b)--(d). The resonance terms use $L=0$ in (a,b), $L=1$ in (c), and $L=2$ in (d). Grey dotted, red solid, and purple dash-dotted curves show the fixed vector baseline, total fit, and radiative contribution. The total and radiative curves include the fitted normalization. Error bars include statistical and systematic uncertainties.}
\label{fig:smbaryon}
\label{fig:smmesons}
\end{figure*}

Panels (a) and (b) of Fig.~\ref{fig:smbaryon} compare the $\Lc$ fits with fixed and free mass and width. We adopt the fixed input $(M_R,\Gamma_R)=(4629,24.2)\MeV$ from the previous analysis of the $B^0\to K_S^0\Lc$ line shape~\cite{LiuEta2026}. Both fixed inputs lie within the corresponding individual $1\sigma$ intervals of the free fit. The local significance with fixed mass and width is $3.2\sigma$.

Panels (c) and (d) of Fig.~\ref{fig:smmesons} show the meson fits. With the masses and widths fixed at their fitted values, the local significances are $3.8\sigma$ for $D_sD_{s1}$ and $2.9\sigma$ for $D_sD_{s2}^{*}$. Fixing the mass and width in $D_sD_{s2}^{*}$ to $(4629,24.2)\MeV$ gives $\chi^2/\mathrm{dof}=9.69/19$ and a local significance of $2.7\sigma$; both inputs lie within the individual $1\sigma$ intervals of the free fit.

\begin{widetext}
\noindent\begin{minipage}{\textwidth}
\centering
\captionof{table}{Fit results for the three Belle spectra with fixed vector baselines. The vector baselines alone give $\chi^2/\mathrm{dof}=17.12/16$, $35.72/22$, and $17.12/20$ in these channels, respectively. Radiative products use the same Born normalization as the fitted spectra and the mass intervals given above. The coefficient $C_i$ is the fitted amplitude normalization for channel $i$. Errors correspond to $\Delta\chi^2=1$.}
\label{tab:smfits}
\small
\setlength{\tabcolsep}{3pt}
\renewcommand{\arraystretch}{1.12}
\begin{tabular*}{\textwidth}[t]{@{\extracolsep{\fill}}lcccc}
\hline\hline
 & \multicolumn{2}{c}{$\Lc$} & $D_sD_{s1}$ & $D_sD_{s2}^{*}$\\
 & Fixed $M_R,\Gamma_R$ & Free $M_R,\Gamma_R$ & Free $M_R,\Gamma_R$ & Free $M_R,\Gamma_R$\\
\hline
$M_R$ [MeV] & 4629 & $4640.1^{+6.0}_{-13.0}$ & $4627.3^{+6.7}_{-7.0}$ & $4625.1^{+20.3}_{-9.5}$\\
$\Gamma_R$ [MeV] & 24.2 & $28.7^{+33.8}_{-23.2}$ & $23.9^{+16.2}_{-15.2}$ & $35.9^{+66.7}_{-28.3}$\\
$C_i$ [$(\mathrm{pb}/\mathrm{GeV})^{1/2}$] & $1220.1^{+185.5}_{-212.6}$ & $1181.4^{+1017.9}_{-300.7}$ & $748.9^{+346.8}_{-164.0}$ & $776.5^{+606.4}_{-263.3}$\\
$\sigma_R(M_R^2)$ [pb] & $435.7^{+142.6}_{-138.6}$ & $447.1^{+1105.3}_{-205.6}$ & $241.2^{+277.2}_{-93.3}$ & $218.4^{+451.4}_{-115.9}$\\
$\sigma_i^{\rm rad}$ [fb] & $76.9^{+25.2}_{-24.5}$ & $93.4^{+40.1}_{-34.8}$ & $42.9^{+14.1}_{-13.4}$ & $60.7^{+29.6}_{-26.6}$\\
$\chi^2/\mathrm{dof}$ & $6.81/15$ & $4.67/13$ & $21.61/19$ & $8.87/17$
\\\hline\hline
\end{tabular*}
\end{minipage}
\end{widetext}

\section{Spectroscopy calculations}
\label{sm:screened}

We calculate the charmonium masses and wave functions in the modified Godfrey--Isgur (MGI) model used in our previous study~\cite{LiuEta2026}. The linear scalar confinement $br+c$ is replaced by a screened form, written before smearing as
\begin{equation}
V_{\rm conf}(r)=c+\frac{b}{\mu}\left(1-e^{-\mu r}\right),
\label{eq:smscreening}
\end{equation}
Here $r$ is the quark--antiquark separation, and $\mu$ is the screening parameter. The color Coulomb and spin dependent interactions, Gaussian smearing, and momentum dependent relativistic corrections follow the GI formalism~\cite{GodfreyIsgur} as used in Refs.~\cite{WangHigherCharm2020,LiuEta2026}.

We adopt the central parameters $m_c=1.65\GeV$, $\mu=0.12\GeV$, $b=0.238\GeV^2$, and $c=-0.337\GeV$ from Ref.~\cite{LiuEta2026}. Following the same study, we estimate the mass uncertainties by varying $\mu=0.12\pm0.01\GeV$. The values $(b,c)=(0.230\GeV^2,-0.329\GeV)$ and $(0.246\GeV^2,-0.344\GeV)$ are used at $\mu=0.11$ and $0.13\GeV$, respectively, with all other parameters unchanged. The quoted errors give the shifts from the central masses to the largest and smallest masses obtained with these three parameter sets.

Table~\ref{tab:smmasses} gives the calculated masses and model uncertainties of the nearby pseudoscalar and $\chi_{cJ}$ states. The central masses of the $\chi_{cJ}(5P)$ and $\chi_{cJ}(6P)$ multiplets lie below and above the fitted meson masses, while $\eta_c(6S)$ lies close to the enhancement region.

\begin{table}[!htbp]
\caption{Calculated masses in MeV. The errors show the model variation under the three parameter sets specified in the text.}
\label{tab:smmasses}
\centering
\setlength{\tabcolsep}{4pt}
\renewcommand{\arraystretch}{1.12}
\begin{ruledtabular}
\begin{tabular}{lrlr}
State & Mass & State & Mass\\
$\eta_c(6S)$ & $4606.6^{+24.4}_{-24.9}$ & $\chi_{c2}(5F)$ & $4726.4^{+33.7}_{-34.7}$ \\
$\chi_{c0}(5P)$ & $4552.4^{+19.7}_{-20.1}$ & $\chi_{c0}(6P)$ & $4678.1^{+30.6}_{-31.3}$\\
$\chi_{c1}(5P)$ & $4563.1^{+20.4}_{-20.8}$ & $\chi_{c1}(6P)$ & $4685.9^{+31.3}_{-32.0}$\\
$\chi_{c2}(5P)$ & $4572.4^{+21.0}_{-21.5}$ & $\chi_{c2}(6P)$ & $4693.0^{+32.0}_{-32.8}$ 
\end{tabular}
\end{ruledtabular}
\end{table}

The strong decays are calculated in the quark pair-creation (QPC) model~\cite{Micu,LeYaouanc}, in which a light $q\bar q$ pair is created with $J^{PC}=0^{++}$. The strengths for creating $u\bar u/d\bar d$ and $s\bar s$ pairs are $\gamma_{u,d}=5.84$ and $\gamma_s=5.84/\sqrt3$, respectively, and the constituent quark masses are $m_{u,d}=0.22\GeV$ and $m_s=0.55\GeV$. The wave functions of the final mesons are taken from Refs.~\cite{Song:2015D,Song:2015Ds}. The final meson masses are $(m_{D_s},m_{D_{s1}},m_{D_{s2}^{*}})=(1.970,2.536,2.573)\GeV$.

For $D_{s1}(2536)$, we use the spin mixture $|D_{s1}(2536)\rangle=\cos\phi\,|{}^3P_1\rangle-\sin\phi\,|{}^1P_1\rangle$, with $\phi=-54.7^\circ$. The decay calculations use the $D_sD_{s2}^{*}$ fit mass, $4625.1\MeV$, for $\eta_c(6S)$. For the candidate formed by mixing $\chi_{c2}(6P)$ and $\chi_{c2}(5F)$, we use the $D_sD_{s1}$ fit mass of $4627.3\MeV$. The partial widths are calculated with the amplitude normalization of Ref.~\cite{LiuEta2026},
\begin{equation}
\Gamma_i(M_R)=\frac{\pi}{4}\frac{q_i}{M_R^2}\sum_{J,L}\left|\mathcal M_i^{JL}(q_i)\right|^2.
\label{eq:smwidth}
\end{equation}
Here $\mathcal M_i^{JL}$ is the QPC partial-wave decay amplitude for channel $i$ with final total spin $J$ and relative orbital angular momentum $L$. The two final mesons have momentum magnitude $q_i$ in the resonance rest frame. Charge conjugates are included.

To illustrate possible mixing of $\chi_{c2}(6P)$ and $\chi_{c2}(5F)$, we denote the lower and higher mass states by $R_L$ and $R_H$, respectively:
\begin{align}
|R_L\rangle&=\cos\theta\,|6P\rangle-\sin\theta\,|5F\rangle,\nonumber\\
|R_H\rangle&=\sin\theta\,|6P\rangle+\cos\theta\,|5F\rangle,
\label{eq:smmixing}
\end{align}
Here $|6P\rangle$ and $|5F\rangle$ denote the unmixed $\chi_{c2}(6P)$ and $\chi_{c2}(5F)$ states, both with $J^{PC}=2^{++}$, and $\theta$ is the mixing angle. Using the central $6P$ and $5F$ masses, we choose $|\theta|=39.15^\circ$ to place $R_L$ at the fitted $D_sD_{s1}$ mass of $4627.3\MeV$. This angle illustrates possible additional mixing.

\noindent\begin{minipage}{\columnwidth}
\captionof{table}{Partial widths in keV, using the $D_sD_{s2}^{*}$ fitted mass for $\eta_c$ and the $D_sD_{s1}$ fitted mass for $R_L$.}
\label{tab:smdecays}
\centering
\setlength{\tabcolsep}{3pt}
\renewcommand{\arraystretch}{1.12}
\begin{tabular*}{\columnwidth}{@{\extracolsep{\fill}}lrcc}
\hline\hline
State & $M_R$ [MeV] & $D_sD_{s1}$ & $D_sD_{s2}^{*}$\\
\hline
$\eta_c(6S)$ & 4625.1 & 0 & 1.4\\
$R_L(+39.15^\circ)$ & 4627.3 & 4.2 & 0.1\\
$R_L(-39.15^\circ)$ & 4627.3 & 5.1 & 0.1\\
\hline\hline
\end{tabular*}
\end{minipage}\par\medskip

Table~\ref{tab:smdecays} shows that the calculated $\eta_c(6S)$ partial width into $D_sD_{s2}^{*}$ is larger than those of the mixed $\chi_{c2}$ candidates. We therefore use the $\eta_c(6S)$ candidate for this channel and a tensor candidate for $D_sD_{s1}$, where the $\eta_c$ decay is forbidden. Additional tensor contributions to $D_sD_{s2}^{*}$ remain possible; their relative yields depend on radiative production rates, full resonance widths, and reconstruction efficiencies. Further angular measurements can help separate the contributions.

\bibliography{references}
\end{document}